\documentclass[aps,prb,showpacs,twocolumn,amssymb,floats,epsfig]{revtex4}

\usepackage{graphicx}
\usepackage{epsfig}
\usepackage{amsfonts}
\usepackage{amsmath}
\usepackage{amssymb}
\usepackage{subfigure}
\usepackage{color}

\newcommand\ba{\begin{eqnarray}}
\newcommand\ea{\end{eqnarray}}
\newcommand\be{\begin{equation}}
\newcommand\ee{\end{equation}}
\newcommand\bi{\bibitem}
\newcommand{\ct}{\cite}
\newcommand{\bra}[1]{\langle #1|}
\newcommand{\ket}[1]{|#1\rangle}

\def\non{\nonumber}

\begin{document}

\title{A study of excess energy and decoherence factor of a qubit coupled to a one dimensional periodically driven spin chain }
\author{Tanay Nag }
\affiliation{\small{
Department of Physics, Indian Institute of Technology, Kanpur 208 016, 
India}}

\begin{abstract}

We take a central spin model (CSM), consisting of a one dimensional environmental Ising spin chain
and a single qubit connected globally to all the spins of the environment, to study numerically the excess energy (EE) of the environment and 
the logarithm of decoherence factor namely, dynamical fidelity susceptibility per site (DFSS),
associated with the qubit under a periodic driving of the transverse field term of environment across its critical point
using the Floquet technique. 
The  coupling to the qubit, prepared in a pure state, with the transverse field
of the spin chain yields  two sets of EE corresponding to the two species of Floquet operators.
In the limit of weak coupling,
we derive an approximated expression of DFSS after an infinite number of driving period which can successfully 
estimate the low and intermediate frequency behavior of numerically obtained DFSS. Our main focus is to 
 analytically investigate the effect of system-environment coupling strength on the  EEs and DFSS and relate the behavior of 
 DFSS to EEs as a function of frequency by plausible analytical arguments. 
We explicitly show that the 
low-frequency beating like pattern of DFSS is an outcome of two  frequencies, causing the oscillations in the two branches of
EEs, that 
are dependent on the coupling strength. In the intermediate frequency
regime, dip structure  observed in DFSS can be justified by the resonance peaks of EEs   
at those coupling parameter dependent frequencies; high frequency saturation value of EEs and DFSS are also connected 
to each other by the coupling strength. 
\end{abstract}

\maketitle

\section{Introduction}
\label{intro}
Dynamics of a periodically driven closed quantum system has been studied extensively in recent years. 
These studies are truly interdisciplinary in nature and are being carried out from the viewpoint of quenching dynamics of
many body quantum systems \ct{zurek05,polkovnikov05,damski05,mukherjee07,mukherjee09,dziarmaga10,polkovnikov11} and the quantum information 
theory\ct{barankov08,sengupta09,quan10,nag11,nag12a,mukherjee12,suzuki15}. 
The behavior of these quantum systems are very much different when compared to the  classical
periodically driven systems \ct{shirley65,grossmann91,hanggi98,kayanuma94,canovi12,russomanno12}.
Many interesting phenomena like coherent destruction of tunneling\ct{das10}, a periodic steady state behavior
of various thermodynamic observables\ct{russomanno12,bastidas12,russomanno13} and  dynamical localization\ct{lazarides14,nag14,nag15}, etc.,  manifest in a  
periodically driven quantum system. A lot of attention has been paid to  Floquet technique due to its successful execution in many periodically driven systems such as
Floquet Graphene\ct{gu11,kitagawa11}, Floquet topological insulator\ct{gomez12}; some of them have also been realized experimentally \ct{kitagawa12}.
Moreover, the excess energy (EE) and the work distribution function for the transverse Ising model
 subjected to a time periodic variation of transverse field have also been studied extensively using Floquet technique \ct{russomanno15}.

Given the recent interest in quenching dynamics of quantum systems, there has been a plethora of studies connecting quantum information theory\ct{nielsen00,vedral07} to
the quantum critical system\ct{chakrabarti96,sachdev99,dutta15}.
The loss of coherence in a quantum system due to its interaction with the environment namely, decoherence, quantified as
decoherence factor (DF) also known as  Loschmidt echo,
is an emerging topic that has been studied greatly in this regard\ct{zurek03,quan06}. The quenching 
dynamics of the DF\ct{damski11a,nag12b,roy13,sachdeva14} has also been investigated throughly for
integrable as well as non-integrable  quantum system under the scope of a central
spin model (CSM)\ct{CSM1,cucchietti07,venuti10,suzuki15}. 
Importantly, in this connection, it has been shown that decohered density matrix of a quantum system under consideration might not always result in the accurate  behavior of 
dynamical fidelity susceptibility per site (DFSS), defined through the logarithm of squared modulus overlap between the initial state
and the time-evolved final state, in the infinite time limit while the quantum system is periodically driven across its QCP\ct{sharma14}. 
Furthermore, there is an experimental observation of quantum criticality has been made by investigating the behavior of
Loschmidt echo, measured without associating an external qubit, in an antiferromagnetic Ising spin chain with finite number of spins using NMR quantum simulator \ct{zhang09}.

We shall consider here a CSM with an environment as 1-d transverse Ising spin chain and a single qubit, weakly and globally coupled to the transverse field term of the chain, to investigate the non trivial 
effect caused by the coupling parameter on the behavior of
the EE of that periodically driven environment across its QCP. Consequently, this coupling leads to two channels of time evolution
for the environment with the modified transverse fields.  These phenomena allows us to exclusively probe the decohering phenomena of the qubit by examining the 
DFSS as a function of frequency. In contrast to the Ref. [\onlinecite{sharma14}] where the overall loss of phase coherence of a periodically driven spin chain 
across its QCP has been studied using 
the Floquet technique, our main focus is to probe the effect of the coupling parameter on DFSS, quantified through the ratio of logarithm of the modulus
squared overlap between two states reached after an infinite number of period of time evolution
with two environmental channels and system size of the environment, of the external qubit under a sinusoidal driving of the transverse field across the QCP.
To the best of our knowledge, this is the first work to examine the behavior of the DFSS of a qubit using the Floquet technique under the 
scope of CSM. Most interestingly,  we show that the behavior of DFSS can be characterized by analyzing the behavior of EEs associated to the two evolution channels 
of the spin chain environment in different frequency regimes.

% In contrast to the Ref. [\onlinecite{sharma14}], we shall consider here a CSM with an environment as 1-d transverse Ising spin chain and a single qubit, coupled to the transverse field of the chain, to investigate the behavior of
% the EE of that periodically driven environment across its QCP. Consequently, the coupling to the qubit leads to two channels of time evolution
% for the environment with the modified transverse fields. These phenomena allows us to probe the decohering phenomena of the qubit by investigating 
% DFSS as a function of frequency. It is quantified through the ratio of logarithm of the modulus squared overlap between two states reached after an infinite number of period of time evolution
% with two environmenal channels and system size of the environment. 
% Our main focus is to probe the effect of the coupling strength on the behavior of EEs associated with two channels. 
% Most interestingly, the DFSS can be charaterised by analysing the behavior of EEs with frequency. 

The paper is organized in the following way. In Sec.~(\ref{model}), we introduce the transverse Ising spin chain and the corresponding
central spin model where a single 
qubit is considered as the system with the spin chain acting as an environment. In parallel, we  present Floquet technique
and express the wave-function for a  periodically driven system in the Floquet representation.  We also elaborately discuss the Floquet machinery in our case 
to probe the EEs of the  environment and the DFSS of the system. In Sec.~(\ref{result}), We analyze the behavior of quasi-energy, EE and DFSS, obtained numerically, 
by explicit analytical calculations with the plausible argument.  Finally, we present concluding comments in Sec.~(\ref{remarks}).

\section{Model and Floquet Technique}
\label{model}

The Hamiltonian $H_E$ of the environment is the ferromagnetic Ising
spin chain in a transverse field consisting of $N$ spins given by \ct{bunder99}
\be
H_E=-\sum_{i=1}^N [\sigma_i^x\sigma_{i+1}^x +h \sigma_i^z], 
\label{eq_ising}
\ee
where $\sigma_x$ and $\sigma_z$ are the Pauli matrices. This model can be exactly solved by
mapping the spins to spinless Fermion through Jordon-Wigner transformation \ct{bunder99,lieb61}. The model
can be decomposed to $2\times 2$ Hamiltonian in the momentum space under periodic boundary condition. The momentum space Hamiltonian
is given by,
\be{\cal H}^E_k = \left[ \begin{array}{cc} 
-h(t)+\cos k & \sin k \\
\sin k  & h(t)-\cos k \end{array} \right], \label{eq:ham_old} \ee
The model has 
 quantum phase transitions (QPTs) at $h=\pm 1$. This QCP belongs to Ising universality class
with $\nu=1$ and $z=1$, where $\nu$ is the correlation length exponent and $z$ is the 
dynamical exponent. This model has a ferromagnetic phase for $|h|< 1$ and a 
paramagnetic phase for $|h|> 1$. In our case, the transverse field is being subjected to a time
periodic sinusoidal driving: $h(t)=1+h_0 \cos(\omega t)$, where $h_0$ is the amplitude
of the driving and $\omega=2\pi/T$ is the frequency of the driving with time period $T$. In our ramp protocol,  
spin chain experiences the QPT only at $h=1$ as the transverse field $h$ is varied between $2$ and $0$.

Let us now discuss the central spin model in which a single spin-$1/2$ particle (qubit) is globally connected to all the spins of the 
environmental spin chain with an interaction Hamiltonian $H_{SE}=\delta\sum_{i=1}^N\sigma_i^z\sigma_S^z$,
where $\sigma_i^z$ is the $i-$th spin of the $XY$ chain, $\sigma_S^z$ represents that of the
qubit and $\delta$ is the coupling strength. 
%Here, we consider that the coupling is switched on simultaneously with the periodic driving of the transverse field. 
Here, we consider that the transverse field initially at $t=0^-$ is $h=1+h_0$ and the coupling to the qubit is suddenly made at $t=0^+$ and simultaneously 
the sinusoidal driving is being started.

%As a result, the ubsequently  environmental time evolution is carried with the above modified transverse field namely, 
%$h(t>0)\pm \delta$.

We choose the qubit to be initially in a pure state at $t =0$, 
 $|\phi_S(t=0)\rangle= a_+|\uparrow\rangle+a_-|\downarrow\rangle$, where $|\uparrow\rangle$ and 
$|\downarrow \rangle$ represent up and down states of the qubit, respectively, and the
environment is in the ground state $|\phi_E (t=0)\rangle = |\phi_g \rangle$. 
The ground state of the composite Hamiltonian $H_E + H_{SE}$, at $t=0$, is then given by their direct product
\be
|\psi(t=0)\rangle=|\phi_S(t=0)\rangle \otimes |\phi_g\rangle.
\label{eq_state_t0}
\ee
It can be shown that at a later  time $t$, the composite wave-function is given by \ct{damski11a}
\be
|\psi(t)\rangle 
=a_+|\uparrow\rangle \otimes |\phi(+,t)\rangle + a_-|\downarrow \rangle \otimes |\phi(-,t)\rangle,
\label{eq_state_t}
\ee
where $|\phi(\pm,t)\rangle$ are the environmental wave-functions, 
satisfying the time dependent schr\"odinger equation: $i|\dot \phi (\pm,t)\rangle=H_E(h(t)\pm\delta)|\phi(\pm,t)\rangle$, evolved from the 
initial ground state wave-function $|\phi_g(t=0^-)\rangle$ before the qubit gets coupled to the environment.
The form of the interaction Hamiltonian generates two evolution channels with 
the modified transverse field as $(h(t)\pm \delta)$. 
Therefore, the modified Hamiltonians ${\mathcal{H}}^E_k(\pm,t)$ which govern the time evolution of environmental spin chain
in momentum space is given by
\be \mathcal{H}^E_k(\pm,t) = \left[ \begin{array}{cc} 
-h(t)\mp \delta+\cos k & \sin k \\
\sin k  & h(t)\pm\delta-\cos k \end{array} \right], \label{eq:ham_new} \ee

We can now focus on  the Floquet theory for a generic time-periodic Hamiltonian, 
$H\left(t\right)=H\left(t+T\right)$. One can construct a time evolution operator for a single period which is 
referred as the Floquet operator 
${\cal F}={\mathcal{O}}e^{-i\int_0^T H(t) dt}$, where $\mathcal{O}$ denotes
time-ordering. $\hbar$ is set to unity. The solution of the Schr\"odinger equation for the $j$-th 
state in the Floquet basis ($\ket{\eta_j(t)}$ which are eigenstates of 
${\cal F}$) can be written in the form $\ket{\Psi_j(t)}=e^{-i\mu_j t}
\ket{\eta_j(t)}$. The states $\ket{\eta_j(t)}$'s 
are time periodic ($\ket{\eta_j(t)}=\ket{\eta_j(t+T)}$) and $e^{-i\mu_j T}$
are the corresponding eigenvalues of ${\cal F}$; the $\mu_j$'s are called 
Floquet quasi-energies. Now, making use the fact that the environmental spin
chain reduces to $2 \times 2$ momentum space Hamiltonian, one can construct a momentum space Floquet
operator ${\mathcal F}_k$ at the stroboscopic time $t=T$. For a sinusoidally varying parameter, one has to
numerically find out the Floquet operator starting from a generic state $(0~1)^{\mathcal T}$, $\mathcal T$ denotes the transpose of a matrix. 
The time evolved wave-function at $t=T$ is given by $(u_k~v_k)^{\mathcal T}$.
Therefore, the ${\mathcal F}_k$ can be constructed from the above state, satisfying the constraint that ${\mathcal F}_k(t=0)$ is an identity, is given by 
\be \mathcal{F}_k = \left[ \begin{array}{cc} 
u_k & -v_k^* \\
v_k & u_k^* \end{array} \right], \label{eq:specf} \ee

Now in our case, one can get two 
Floquet operators $\mathcal{F}_k(\pm)$ for two channels of evolution associated with the 
modified transverse fields $h(t)\pm \delta$. By diagonalizing the Floquet
operators one can get $\mu_k^{\pm}(\pm)$, quasi-energies and $\ket{\eta_k^{\pm}(\pm)}$, quasi-states corresponding 
to two channels of evolution. ($\mathcal{F}_k(+)$ gives two quasi-states $\ket{\eta_k^{\pm}(+)}$, same as for the
negative channel).  Under the periodic driving, the time evolved environmental state $|\phi_k(\pm)\rangle$ at time 
$t=nT$, can be obtained  in terms of Floquet basis sates,
\ba
\ket{\phi_k(\pm,nT)}&=&c_k^+(\pm) \mathrm{e}^{-i\mu_k^+(\pm) \,nT} \ket{\eta_k^+(\pm)} \nonumber \\
&+& c_k^-(\pm) 
\mathrm{e}^{-i\mu_k^-(\pm) \,nT} \ket{\eta_k^-(\pm)},
\label{eq_state_fl}
\ea
where $c_k^{\pm}(\pm)=\langle 
\eta_k^\pm(\pm) \ket{\phi_{g,k}}$, with $\ket{\phi_{g,k}}$ is the initial bare ground 
state of the environment for a momentum mode $k$  at $t=0^-$ when the qubit is not coupled with the environment.

We can now probe the EE and DFSS as a function of frequency by using environmental wave-function. We have two 
 sets of EE, associated with the two channels of time evolution, are given by  
\ba 
W(\pm,nT) &=& (1/N) \sum_k W_k(\pm,nT) \equiv (1/N) \sum_k [e_k(\pm,nT)\nonumber \\
&-& e_{g,k}(\pm,nT)],
\label{eq_WD}
\ea
where $e_k(\pm,nT)$ is the energy expectation value of the environmental Hamiltonian (\ref{eq:ham_new}) for the $k$-th mode, reached after $n$-th 
time period,  given by $e_k(\pm,nT)= \bra{\phi_k(\pm,nT)} \mathcal{H}^E_k(\pm,nT) \ket{\phi_k(\pm,nT)}$; $e_{g,k}(\pm,nT)$ is the 
ground state energy of the ${\mathcal{H}}^E_k(\pm,nT)$. 

In the limit of $n\to\infty$, we resort to the Riemann-Lesbesgue lemma for integrating out the rapidly oscillating phase factor. 
The EEs for the two channels by retaining only the steady state contribution in infinite time limit are given by 
\ba
W(\pm)&=&(1/N)\sum_k\biggl[|c_k^{+}(\pm)|^2\langle \eta_k^+(\pm)|H(\pm,T)|\eta_k^+(\pm)\rangle \non \\
&+& |c_k^{-}(\pm)|^2 \langle \eta_k^-(\pm)|H(\pm,T)|\eta_k^-(\pm)\rangle \non \\
&-&e_{g,k}(\pm,T)\biggr] 
\label{eq_WDI}
\ea

We shall now find out the DF of the qubit from the reduced density matrix
of the qubit by tracing over environmental part from the composite density matrix. The reduced density matrix in the
$\protect{\{|\uparrow\rangle,|\downarrow\rangle\}}$ basis reads as
\be \mathcal{\rho}_S(t)= {\rm Tr}_{\cal E}\biggl[|\psi(t)\rangle\langle\psi(t)|\biggr]=
\left[ \begin{array}{cc} 
|a_+|^{2} &   a_+ a_-^* d^*(t) \\
a_+^* a_- d(t) & |a_-|^{2} \end{array} \right], \label{eq:dm} \ee
where $d(t)=\langle\phi_+(t)|\phi_-(t)\rangle$ appears as an off-diagonal 
element in the reduced density matrix and is related to the DF $D(t)$ by a modulo square of $d(t)$\ct{damski11a}, 
$D(t)=|\langle \phi_+(t)|\phi_-(t)\rangle|^2$.  DF measures the purity of the qubit; $D=1$ signifies that the 
qubit is in a pure state.

In the momentum space language, $D(t)$ is given by 
$D(t)=\Pi_{k>0}|d_k(t)|^2=\Pi_{k>0}|\langle\phi_k(+,t)|\phi_k(-,t)\rangle|^2$.
%In the infinite time limit ($n\to \infty$), one can neglect the 
%highly oscillating terms consisting of $\exp(i\mu_k^{\pm}(\pm)nT)$ ,
Similarly, one can measure the DF of qubit after $n$-th cycle of the time periodic transverse field. 
Therefore, the DF  in its rudimentary form is therefore given by
\ba
&D(n)&=\Pi_{k>0}D_k(n)\nonumber\\
&=&\Pi_{k>0} \biggl[|c_k^+(+)|^2|c_k^+(-)|^2 ~|\langle\eta_k^+(+)|\eta_k^+(-)\rangle|^2\nonumber\\
&+&|c_k^-(+)|^2|c_k^-(-)|^2 ~|\langle\eta_k^-(+)|\eta_k^-(-)\rangle|^2\nonumber\\
&+& |c_k^+(+)|^2|c_k^-(-)|^2|\langle\eta_k^+(+)|\eta_k^-(-)\rangle|^2\nonumber\\
&+&|c_k^-(+)|^2|c_k^+(-)|^2 |\langle\eta_k^-(+)|\eta_k^+(-)\rangle|^2\nonumber\\
&+& 2~ {\rm Re}\biggl[T'_1+T'_2+T'_3+T'_4+T'_5+T'_6\biggr] \biggr],
\label{eq_df}
\ea
where the $T_m$'s denote the cross terms which come in pairs. Here, we use $T'_m+(T'_m)^*=2~{\rm Re}[T'_m]$. 
%with exponential factor consisting of $\exp(i\mu_k^{\pm}(\pm)nT)$.
For our convenience, we 
shall segregate the exponential part from the $T'_m$ i.e.,
$T'_m=T_m\times {\rm exp}(i \alpha_m)$.  $T_m$'s and $\alpha_m$'s are given by

\ba
T_1&=& c_k^+(+)^*c_k^+(-)c_k^-(+)c_k^-(-)^* \langle\eta_k^+(+)|\eta_k^+(-)\rangle \nonumber\\
&\times& \langle\eta_k^-(-)|\eta_k^-(+)\rangle, \nonumber\\
T_2&=& |c_k^+(+)|^2 c_k^-(-)^*c_k^+(-) \langle\eta_k^+(+)|\eta_k^+(-)\rangle \langle\eta_k^-(-)|\eta_k^+(+)\rangle, \nonumber\\
T_3&=&|c_k^+(-)|^2 (c_k^+(+))^*c_k^-(+) \langle\eta_k^+(+)|\eta_k^+(-)\rangle \langle\eta_k^+(-)|\eta_k^-(+)\rangle,\nonumber\\
T_4&=&|c_k^-(-)|^2c_k^-(+)^*c_k^+(+) \langle\eta_k^-(+)|\eta_k^-(-)\rangle \langle\eta_k^-(-)|\eta_k^+(+)\rangle,\nonumber\\
T_5&=&|c_k^-(+)|^2c_k^+(-)^*c_k^-(-) \langle\eta_k^-(+)|\eta_k^-(-)\rangle \langle\eta_k^+(-)|\eta_k^-(+)\rangle,\nonumber\\
T_6&=&c_k^+(+)^* c_k^-(-) c_k^-(+) c_k^+(-)^* \langle\eta_k^+(+)|\eta_k^-(-)\rangle\nonumber\\
&\times&  \langle\eta_k^+(-)|\eta_k^-(+)\rangle, \nonumber\\
\alpha_1&=&2nT(\mu_k^+(+)-\mu_k^+(-)),~\alpha_2=-2nT\mu_k^+(-),\nonumber\\
\alpha_3&=&2nT\mu_k^+(+),~\alpha_4= -2nT\mu_k^+(+),\nonumber\\
\alpha_5&=&-2nT\mu_k^-(-),~\alpha_6=2nT(\mu_k^+(+)-\mu_k^-(-)) \non 
\label{eq_df2}
\ea

% &+&2~{\rm Re}\biggl((c_k^+(+))^*c_k^+(-)c_k^-(+)(c_k^-(-))^*e^{2inT\mu_k^+(+)}\nonumber\\
% &\times&e^{-2inT\mu_k^+(-)}\langle\eta_k^+(+)|\eta_k^+(-)\rangle \langle\eta_k^-(-)|\eta_k^-(+)\rangle\biggr) \nonumber\\

% &+&2~{\rm Re}\biggl(|c_k^+(+)|^2c_k^+(-)(c_k^-(-))^*e^{-2inT\mu_k^+(-)}\nonumber\\
% &\times&\langle\eta_k^+(+)|\eta_k^+(-)\rangle \langle\eta_k^-(-)|\eta_k^+(+)\rangle\biggr)

%+2~{\rm Re}\biggl(|c_k^+(-)|^2 \nonumber\\ 
% &\times& (c_k^+(+))^*c_k^-(+)e^{2inT\mu_k^+(+)} \langle\eta_k^+(+)|\eta_k^+(-)\rangle \nonumber\\
% &\times&\langle\eta_k^+(-)|\eta_k^-(+)\rangle\biggr)

%+2~{\rm Re}\biggl( |c_k^-(-)|^2(c_k^-(+))^*c_k^+(+) \nonumber\\ 
% &\times& e^{-2inT\mu_k^+(+)}\langle\eta_k^-(+)|\eta_k^-(-)\rangle \langle\eta_k^-(-)|\eta_k^+(+)\rangle\biggr)\nonumber\\

% &+&2{\rm Re}\biggl( |c_k^-(+)|^2(c_k^+(-))^*c_k^-(-) e^{-2inT\mu_k^-(-)} \langle\eta_k^-(+)|\eta_k^-(-)\rangle       \nonumber\\
% &\times& \langle\eta_k^+(-)|\eta_k^-(+)\rangle  \biggr) 

%+2~{\rm Re}\biggl((c_k^+(+))^* c_k^-(-) c_k^-(+)  \nonumber\\
% &\times&(c_k^+(-))^* e^{2int\mu_k^+(+)} \langle\eta_k^+(+)|\eta_k^-(-)\rangle \langle\eta_k^+(-)|\eta_k^-(+)\rangle \biggr)\biggr] 

Our main aim is to study the behavior of DFSS given by $\chi_F(n\to\infty)=\log(D(n\to\infty))/N=\sum_k {\rm log}(D_k(n\to \infty))/N$ in the infinite time limit. 
 We note the DF of the qubit is appreciably small referring to the 
fact that the qubit is totally in a mixed state after an infinite number of periods and hence, the logarithm of DF namely, fidelity susceptibility is the
main quantity to be studied in this context. The first four terms of the above expression (\ref{eq_df})
give the decohered value which is not the actual value of the DFSS in the $n\to \infty$ limit. In this limit one can not simply
neglect the exponential term, appearing inside the logarithm, of DFSS \ct{sharma14}.

One can consider $T_m$ is a real quantity without loss of generality. Now, in order to simplify the expression in $n\to \infty$ limit
we shall write the $\chi_F(n)$ in the following form
\be
\chi_F(n)
={1 \over N}\sum_k \biggl[\log(D^{\rm dec}_k)+\log\biggl(1+\sum_{m=1}^6 ({x'_m  }\cos\alpha_m)\biggr)\biggr]
\label{eq_df1}
\ee
where $x'_m=2T_m/ D^{\rm dec}_k$, and $D_k^{\rm dec}$ is the decohered part, consisting of first four terms in Eq.~(\ref{eq_df}), for a momentum mode $k$ .
% Here, $x_m$'s and $y_m$'s 
% are the real and imaginary part associated with the $T_m$. $\alpha_m$'s in the expression  (\ref{eq_df1}), containing the product $nT$, are the exponents 
% associated with exponentials present
% in the cross term $T'_m$s of (\ref{eq_df}). $T_m$'s are given by
% \ba
% T_1&=& c_k^+(+)^*c_k^+(-)c_k^-(+)c_k^-(-)^* \langle\eta_k^+(+)|\eta_k^+(-)\rangle \langle\eta_k^-(-)|\eta_k^-(+)\rangle,\nonumber\\
% x_1&=&{\rm Re}[T_1],~y_1={\rm Im}[T_1],\alpha_1=2nT(\mu_k^+(+)-\mu_k^+(-)), \nonumber\\
% T_2&=& |c_k^+(-)|^2 (c_k^+(+))^*c_k^-(+) \langle\eta_k^+(+)|\eta_k^+(-)\rangle, \nonumber\\
% x_2&=&{\rm Re}[T_2],~y_2={\rm Im}[T_2], \alpha_2=-2nT\mu_k^+(-),\nonumber\\
% T_3&=&|c_k^+(-)|^2 (c_k^+(+))^*c_k^-(+) \langle\eta_k^+(+)|\eta_k^+(-)\rangle \langle\eta_k^+(-)|\eta_k^-(+)\rangle,\nonumber\\
% x_3&=&{\rm Re}[T_3],~ y_3={\rm Im}[T_3], \alpha_3=2nT\mu_k^+(+),\nonumber\\
% T_4&=&|c_k^-(-)|^2(c_k^-(+))^*c_k^+(+) \langle\eta_k^-(+)|\eta_k^-(-)\rangle \langle\eta_k^-(-)|\eta_k^+(+)\rangle,\nonumber\\
% x_4&=&{\rm Re}[T_4],~y_4={\rm Im}[T_4],~\alpha_4= -2nT\mu_k^+(+),\nonumber\\
% T_5&=&|c_k^-(+)|^2(c_k^+(-))^*c_k^-(-) \langle\eta_k^-(+)|\eta_k^-(-)\rangle \langle\eta_k^+(-)|\eta_k^-(+)\rangle,\nonumber\\
% x_5&=&{\rm Re}[T_5],~y_5={\rm Im}[T_5],~\alpha_5=-2nT\mu_k^-(-)\nonumber\\
% T_6&=&c_k^+(+)^* c_k^-(-) c_k^-(+) c_k^+(-)^* \langle\eta_k^+(+)|\eta_k^-(-)\rangle \langle\eta_k^+(-)|\eta_k^-(+)\rangle,\nonumber\\
% x_6&=&{\rm Re}[T_6],~y_6={\rm Im}[T_6],~\alpha_6=2nT(\mu_k^+(+)-\mu_k^-(-)).
% \label{eq_df2}
% \ea

Now, in the limit of sufficiently smaller value of $\delta$ and using the asymptotic expansion of logarithmic series $\log(1+x)\simeq x-x^2/2+x^3/3 -\cdots$, one can use the Riemann-
Lesbesgue lemma to achieve the final form of the DFSS in the $n\to \infty$ limit \ct{sharma14}.
The fidelity susceptibility $\chi_F$ is defined as $F(\lambda,\lambda+\delta)=
|\langle\psi_0(\lambda)|\psi_0(\lambda+\delta)\rangle|^2=1-\delta^2 N\chi_F$,  where $\ket{\psi_0(\lambda)}$ is 
the ground state of the quantum system\ct{zanardi06}. Therefore, one can express $\chi_F(n\to\infty)$ up to $O(\delta^2)$
in the power series of $x'_m$'s 
\ba
\chi_F(\infty)&\simeq&\int {dk \over 2\pi}\biggl[ \log D^{\rm dec}_k -\biggl({x'^2_1 \over 4} + {3 x'^4_1 \over 32}+ {5 x'^6_1 \over 96} +\cdots   \nonumber\\
&+& \sum_{m=2}^6 {x'^2_m \over 4} \biggr) +{\rm cross ~terms} \biggr]\nonumber\\
&\approx& \int {dk \over 2\pi}\biggl[ \log D^{\rm dec}_k - \biggl( \log \biggl( {2 \over 1+\sqrt{1-x'^2_1}}\biggr) \nonumber\\
&+&  \sum_{m=2}^5 {x'^2_m \over 4} \biggr) \biggr]
\label{eq_df3}
\ea
The detail of the  above derivation to obtain the simplified and approximated expression 
of $\chi_F(\infty)$ (\ref{eq_df3}) is presented in the Appendix (\ref{ch_append}).

Here, we use the fact that any even multiple of $\cos\alpha_m$ would contribute to the integral 
as they  do  not  average  to  zero when $n$ is  very  large. The cross terms in the $\chi_F(\infty)$ are sum of the product of $x'^{a}_m$ and $x'^{b}_n$ with their 
all possible combinations where $a$ and $b$ are both even numbers and $m\neq n$. 
%One can show that in the small $\delta$ limit the contribution coming from the
%``cross terms" is sub-leading. 
%with respect to that of the logarithmic term consisting of $x'^2_1$. 
% The leading order ``cross terms" are of $O(\delta^4)$ except terms like $x'^a_1x'^2_m$ with $m\neq 1,6$ which has an order $O(\delta^2)$ involved in it; one can not simply 
%replace all the $O(\delta^2)$ correction term originated from the above product by a closed form  expression.
 In the low frequency and intermediate frequency regime the product  terms $x'_1x'_m$ (with $m\neq 1$) are infinitesimally small 
and contribution coming from $x'^a_1x'^2_m$ can be safely neglected; consequently, the $\chi_F(\infty)$ is sufficient
to estimate the behavior exhibited by 
 $\log(D(n))/N$  which is obtained numerically  using a large value of $n$.
Therefore, excluding the above ``cross terms"
we keep up to an $O(\delta^2)$ term, coming from the decohered part, logarithmic part and $x'^2_m$ (with $m=2,~\cdots,~5$),
in calculating $\chi_F(\infty)$ with small but finite $\delta<1$.

One can numerically check that the $x'_1x'_m$ has a significant contribution in
the high frequency regime and $x'^a_1x'^2_m$ can not be neglected.  This might cause a problem for the approximated expression of $\chi_F(\infty)$  (\ref{eq_df3}) in determining
the accurate behavior obtained numerically  by $\chi_F(n)=\log(D(n))/N$ (\ref{eq_df1}) 
with a large value of $n$ in that high frequency limit.
%Furthermore, the summation of these product terms does not simply lead to a closed form 
%expression which is possible for the other non-cross terms involving $x'_m$ and $y'_m$ separately. 
% The   given in Eq.~(\ref{eq_df3})  to expalin the behavior of the DFSS (\ref{eq_df1})in 
% low and intermediate frequency regime. In the high frequency regime, the $O(\delta^2)$ plays the key role to determine the  behavior of DFSS.

% The $x'_m$ and $y'_m$ terms with $m>1$ are in first order in $\delta$ while $x'_1$ and $y'_1$ are having terms like $O(\delta^0)+ O(\delta^2)$. Now,
%  the lowest order of the cross terms, containing $O(x'^2_m x'^2_p, y'^2_m y'^2_p, x'^2_m y'^2_p)$ with all possible combinations of $m$ and $p$, yield
%  $ O(\delta^2,\delta^3)$ contribution to $\chi_F(\infty,\delta)$. Therefore,
%  it can be easily shown that if we are interested in finding an approximated expression of $\chi_F(\infty,\delta)$ with small $\delta<<1$ then
%  the subleading contribution coming from cross terms can safely be neglected even in the lowest order of cross term.
% %When the coupling is not sufficiently small, the $\chi_F$ (\ref{eq_df3}) might not always follow the large $n$ behavior of DFSS (\ref{eq_df1}) as a function of the driving frequency. 
% The detail of the above derivation is presented in the Appendix (\ref{ch_append}). 
Furthermore, one can not simply obtain $\chi_F(\infty,\delta)=0$ by setting $\delta=0$ in $\chi_F(\infty)$ (\ref{eq_df3}). One has to separately treat
the infinitesimally small $\delta \to 0$ case. In order to probe $\delta= 0$ limit, $\chi_F(n)$ (\ref{eq_df1}) is the appropriate 
quantity to begin with. In the case of $\delta=0$, we have only one Floquet operator and as a result 
$|\eta_k^{\pm}(-)\rangle=|\eta_k^{\pm}(+)\rangle=|\eta_k^{\pm}\rangle$, $\mu_k^{\pm}(+)=\mu_k^{\pm}(-)=\mu_k^{\pm}$ and 
$c_k^{\pm}(+)=c_k^{\pm}(-)=c_k^{\pm}$. Therefore, $T'_1$ is the only non-decohered term which contributes to the $\chi_F(n)$ 
(\ref{eq_df1}); all the other cross terms $T'_m$, $m\ne1$, vanish
due the orthogonality condition of quasi-states $\langle \eta_k^{+}|\eta_k^{-}\rangle=0$. It can be easily shown that $\chi_F(n)$ for 
$\delta=0$ reduces to the following form
\ba
\chi_F(n,\delta=0)&=&{1 \over N}\sum_k \log\biggl( |c_k^+|^4 + |c_k^-|^4 + 2 |c_k^+|^2|c_k^-|^2 \biggr)\nonumber\\
&=&{1 \over N}\sum_k \log\biggl( \biggl[|c_k^+|^2+|c_k^-|^2\biggr]^2 \biggr)=0
\label{eq_dfz}
\ea
Since, $\chi_F(n,\delta=0)$ (\ref{eq_dfz}) does not have any sinusoidal term containing $n$ in its argument; this allows us to 
write $\chi_F(n\to \infty,\delta=0)=\chi_F(n,\delta=0)=0$.

Interestingly, the $\chi_F(\infty)$ (\ref{eq_df3}) can not correctly quantify the DFSS when 
$\delta=0$ as the $\log(D_k^{\rm dec})$ and $\log(2/(1+\sqrt{1-x'^2_1}))$ terms of order $O(\delta^0)$ do not cancel each other. In order to 
obtain the $\chi_F(\infty)$ (\ref{eq_df3}), we replace the even power of sinusoidal term by its time averaged value and this results 
in a permanent loss of phase information while calculating $\chi_F(n\to \infty)$. The loss of phase information causes the irreversibility
in the behavior of $\chi_F(\infty)$ i.e., $\chi_F(n)$ can successfully predict the $\delta=0$ behavior where as $\chi_F(\infty)$,
derived from $\chi_F(n)$ with $n\to\infty$, can not correctly quantify the $\delta=0$ behavior. 
%This has been elaborately discussed 
%at the end of the Sec. (\ref{result}).

\section{Results}
\label{result}
\begin{figure}[ht]
\begin{center}
\includegraphics[height=6.0cm]{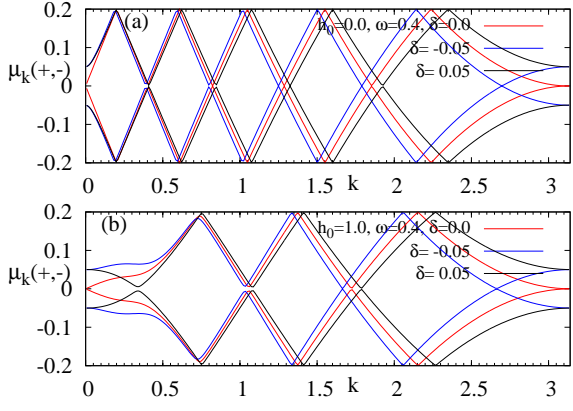}
\end{center}
\caption{Fig. (a) shows the variation of two channels of quasi-energy $\mu_k^{\pm}(\pm)$ as a function of momentum $k$ with
 driving amplitude $h_0=0$. Fig. (b) depicts the behavior of quasi-energy
 while the driving amplitude is finite $h_0=1$.} 
\label{fig_qe}
\end{figure}

In this section, we examine the nature of quasi-energies, EEs and the DFSS in detail.
 We shall first investigate the behavior of the two sectors of quasi-energies associated with the two coupling channels as a function of the momentum.
 The time periodic Hamiltonian causes a temporal Brillouin zone (TBZ) structure of width $\omega$ in 
 the behavior of quasi-energy $\mu_k^{\pm}(\pm)$ and consequently  this originates the quasi-degeneracy in the 
 Floquet spectrum \ct{hanggi98}. Quasi-degeneracy occurs when the one branch of the 
 Floquet spectrum meets with the other branch inside the same TBZ or two adjacent
 TBZs. Investigating each channel of quasi-energy (see Fig.~\ref{fig_qe} (a) and (b)), we find 
 that the coupling strength $\delta$ indeed has an effect on  
 quasi-degenerate momentum mode.

 %The pseudo dispersion of the Hamiltonian (\ref{eq:ham_new}) is given by 
%  \be
%  |E_k(\pm,\omega)|= \sqrt{(-1-h_0 \cos(\omega t)\pm \delta +\cos k)^2+\sin^2 k}.
%  \label{eq_dis}
%  \ee
 In order to find out the quasi-degenerate momentum  value  we have to take the 
 limit $h_0 \to 0$ in Eq.~(\ref{eq:ham_new}).  Therefore, the quasi-degeneracy condition can be obtained by diagonalizing the Hamiltonian (\ref{eq:ham_new}) is given by
 \be
 E_k(\pm)=-E_k(\pm)+l\omega, ~~ l=0,1,2,\cdots
 \label{qe_deg}
 \ee
 where $E_k(\pm)$ is eigenvalue of Hamiltonian (\ref{eq:ham_new}) in the limit $h_0\to 0$.
 Therefore, quasi-degenerate momentum modes $k^{\pm}_q(\delta)\simeq 2\arcsin ((1\pm \delta/2)~l\omega/4)$
 which are quantitatively matching with the quasi-degenerate momentum observed in Fig.~(\ref{fig_qe}a). The point to note is that there is no
 quasi-degeneracy exists for $k=0$ and $\pi$ due to the coupling $\delta$. The $k^{+}_q(\delta) (k^{-}_q(\delta))$ for 
 positive (negative) channel  gets shifted towards right (left) of the $k_q(\delta=0)$.  
This shifting is observed quite prominently as one increases $k$ from $0$ to $\pi$.

 Similarly, we show in Fig.~(\ref{fig_qe}b) that $k^{\pm}_q(\delta)$ shifts towards the Brillouin zone (BZ) boundary
 $k=\pi$ for higher value of $h_0=1$. As a result, less number of 
 quasi-degeneracy appears in Floquet spectrum. Figure~(\ref{fig_wd}) shows that the EEs exhibit peaks at those 
 quasi-degenerate points for finite $h_0$. Hereafter, we shall refer positive channel as $\delta>0$ and negative channel as $\delta<0$ 
 in all the figure.

\begin{figure}[ht]
\begin{center}
\includegraphics[height=6.0cm]{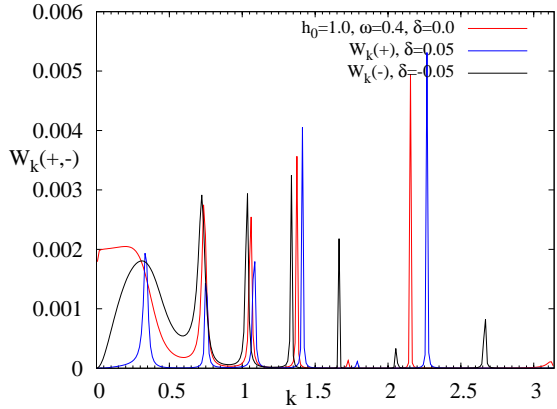}
\end{center}
\caption{Plot shows the variation of EEs $W_k(\pm)$ for two channels with modified ($h(t)\pm\delta$)
 as a function of $k$.} 
\label{fig_wd}
\end{figure}

Now, we shall focus on the behavior of the two channels of EE $W(\pm)$, obtained numerically from Eq.~(\ref{eq_WDI}),
as a function of frequency
$\omega$. Hereafter, we set $h_0=1$. We divide the frequency range into three parts depending upon the distinctive 
behavior of EE in these $\omega$ regimes, (a) low-frequency oscillations of $W(\pm)$, (b) resonance
peak of $W(\pm)$ at the intermediate frequency   and (c) high frequency plateau of EE. 
In all of the above three regions the coupling strength $\delta$ plays an 
important role in determining the behavior of $W(\pm)$.

 Let us first explore the part (a) i.e., the low-frequency regime.  By Investigating Fig.~(\ref{fig_wd}(a) and (b)), one can see that the qualitative behavior of EEs 
for two channels are same, though there are many quantitative differences. The $W(\pm)$ shows 
oscillations while the positions of the minima for each channel are dependent on the coupling strength $\delta$.

In order to probe the low frequency behavior of $W(\pm)$ one has to work with the rotated Hamiltonian and use the perturbation theory.
 The perturbed Hamiltonian near the critical point $h=1$ and close to the critical momentum mode $k=0$ 
looks like ${\cal H}^E_k(\pm)\simeq(-\cos(\omega t)-k^2/2\pm \delta)\sigma_z+k\sigma_x$. Now, we can change the reference frame
using the transformation rule for the environmental wave-function satisfying the Schr\"odinger equation : $\ket{\phi_k(\pm,t)}'=R_k(\pm,t)\ket{\phi_k(\pm,t)}$,
where
\ba
R_k(\pm,t)&=&{\rm exp}[-i\int_0^t(-\cos(\omega t) \pm \delta)~ dt \sigma_z] \nonumber \\
&=&{\rm exp}\biggl[{i(\sin(\omega t)\mp \delta \omega t)\over \omega}\sigma_z\biggr] \nonumber \\
&=& \cos \alpha_{\mp}I +i\sin \alpha_{\mp}\sigma_z, 
\label{R_mat}
\ea
with $\alpha_\mp=(\sin(\omega t)\mp \delta \omega t)/\omega$. Therefore, the modified environmental Hamiltonian in the rotated frame
${\cal H}_k^{'}(\pm)=R_k(\pm,t)~{\cal H}^E_k(\pm)~R_k^{\dagger}(\pm,t)$ is given by
\ba
{\cal H}_k^{'}(\pm,t)&=&
 {1 \over 2}~(-k^2\pm 2\delta - 2\cos(\omega t))\sigma_z  + k \cos (2\alpha_{\mp})\sigma_x \nonumber \\
 &-&2k \sin (2\alpha_{\mp})\sigma_y
\label{hr_mat}
\ea
Now, one has to calculate the time evolution operator $U_k^{'}(\pm)$ in the rotated frame over a single period $T=2\pi/\omega$ using the 
perturbed rotated Hamiltonian (\ref{hr_mat}), $U_k^{'}(\pm,T)={\cal{O}}~{\rm exp}(i\int_0^T {\cal H}_k^{'}(\pm,t) ~ dt)$.
% In order to do that we have to make 
% use the following expansion of Bessel's function
% \ba
% \cos(z\sin\theta)&=&J_0(z)+2 \sum_{n=1}^{\infty}J_{2n}(z)\cos(2n\theta)\nonumber \\
% \sin(z\sin\theta)&=&2 \sum_{n=0}^{\infty}J_{2n+1}(z)\sin((2n+1)\theta) \nonumber \\
% %\label{bessel}
% \ea

\begin{figure}[ht]
\begin{center}
\includegraphics[height=6.0cm]{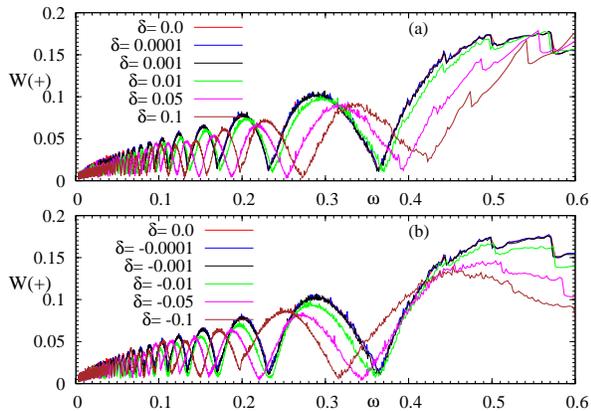}
\end{center}
\caption{Plot shows the variation of EE for two channels with $(h+\delta)$ (in Fig.~(a))
$(h-\delta)$ (in Fig.~(b)) as a function of $\omega$. The minima of $W(-),~(W(+))$ shifts towards left 
(right) of the minima obtained for $W(\delta=0)$.} 
\label{fig_wd}
\end{figure}

Using the properties of Bessel function and by retaining the leading order contribution in the limit $k T\ll 1$ and $\delta < 1$, one can get 
\ba
&U_k^{'}(\pm,t)&\simeq
 I-i\biggl[{1 \over 2}~(-k^2\pm 2\delta)\sigma_z 
 +k \biggl(J_0\biggl({2 \over \omega}\biggr)\nonumber \\
 &\mp&{2\delta~ T \over \pi} J_1\biggl({2 \over \omega}\biggr)\biggr)\sigma_x 
 \mp 2k J_0\biggl({2 \over \omega}\biggr) \delta~ T~\sigma_y \biggr]T \nonumber \\
 &\simeq& I-i\biggl[{1 \over 2}~(-k^2\pm 2\delta)\sigma_z 
 + k \biggl(J_0\biggl({2  \over \omega}\biggr) \nonumber \\
&\mp&{2\delta~ T \over \pi} J_1\biggl({2 \over \omega}\biggr)\biggr)\sigma_x\biggr]T \nonumber \\
&\simeq&I-iT\biggl[{1 \over 2}(-k^2\pm 2\delta)\sigma_z 
 + k J_0\biggl({2  \pm 4 \delta\over \omega}\biggr) \sigma_x\biggr]\nonumber
\label{ur_mat}
\ea  
Now, one can obtain the Floquet operator which is the time evolution operator in the initial frame 
$U_k(\pm,T)=R_k(\pm,T)U'_k(\pm,T)R_k^{\dagger}(\pm,T)$ is given by
\be
U_k(\pm,T)\simeq I -i\biggl[{1 \over 2}~(-k^2\pm 2\delta)\sigma_z 
 \mp 2 k\delta T J_0\biggl({2  \pm 4 \delta\over \omega}\biggr) \sigma_y\biggr]T
\label{u_mat}
\ee

\begin{figure}[ht]
\begin{center}
\includegraphics[height=6.0cm]{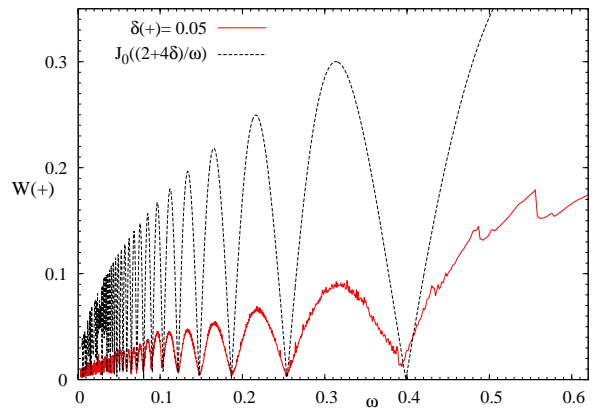}
\end{center}
\caption{Plot shows the minima of EE for positive channel with $(h+\delta)$ are closely
matching with the zeros of the Bessel's function $J_0((2+4\delta)/\omega)$ while both of them are plotted 
as a function of $\omega$. Here, $\delta=0.05$. } 
\label{fig_wd_comp}
\end{figure}

One can therefore estimate the eigenstates (Floquet states) and eigen-energy (quasi-energy) by diagonalizing the Floquet operator (\ref{u_mat}). 
In the low frequency limit $k \ll J_0((2\pm 4 \delta)/\omega)$, the behavior of $W(\pm)$ is  determined  by the behavior of 
Bessel function. In Fig.~(\ref{fig_wd_comp}), we explicitly show that EE for the $(h(t)+\delta)$ channel the 
position of the minima of $W(+)$ matches with the zeros of the modified 
Bessel function $ J_0((2+ 4 \delta)/\omega)$. Therefore, the coupling to the qubit has an effect on the EE of the environmental spin chain.

\begin{figure}[ht]
\begin{center}
\includegraphics[height=6.0cm]{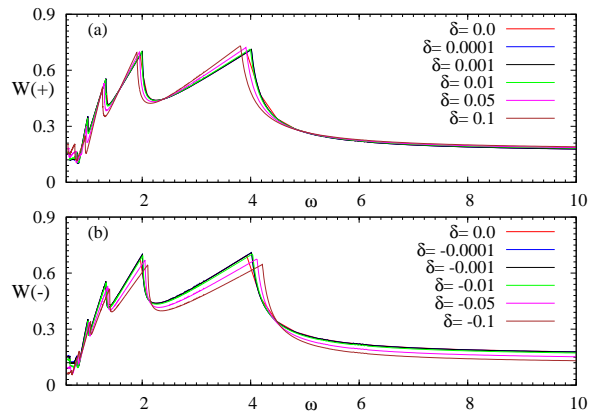}
\end{center}
\caption{Plot shows the variation of EE for two channels with $(h+\delta)$ (in Fig.~(a))
$(h-\delta)$ (in Fig.~(b)) as a function of $\omega$. The peak of $W(-),~(W(+))$ shifts towards right 
(left) of the resonance peaks obtained for $W(\delta=0)$.} 
\label{fig_wd_hf}
\end{figure}

Now, we focus on the response of EE in the intermediate  frequency range where one observes 
a series of resonance peaks (see Fig.~(\ref{fig_wd_hf}(a) and (b)). The position of this resonance peaks are also dependent on the channel of evolution and $\delta$. 
The resonance occurs when the quasi-degenerate momentum obtained from the energy spectrum associated with the Hamiltonian (\ref{eq:ham_new}) 
crosses the edge of the 
BZ $k=\pi$. 
In oder to obtain the position of the peaks once again we have to take the limit $h_0\to 0$. Therefore, the resonance position can be determined from Eq.~(\ref{qe_deg})
and is given by 
\be
\omega_r(\pm)={4\mp2\delta \over l},
\label{eq_reso}
\ee
 where $l=1,2,3, \cdots$. According to this relation the shift of resonance frequency $\omega_r(\pm)$ from the uncoupled case when no qubit
 is coupled to the environmental chain is proportional to $\delta$. For example, the position of peaks for $\delta=0.05$ are
 $\omega_r(+)=3.9, ~1.95,~ 1.3,~\cdots$ (see Fig.~(\ref{fig_wd_hf}a)) which are successfully predicted by the Eq.~(\ref{eq_reso}). This
 resonance peaks are observed until the Bessel function starts dominating the low-frequency behavior of EE. 
 
\begin{figure}[ht]
\begin{center}
\includegraphics[height=6.0cm]{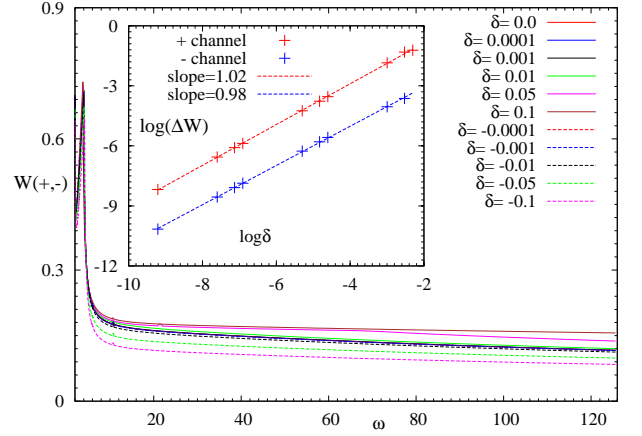}
\end{center}
\caption{Plot shows the variation of EE for the both the channels with $(h\pm\delta)$ in the high $\omega$ limit. 
The saturation value is dependent on $\delta$. Inset shows that the amount of deviation in the saturation value of $W(\pm)$
from that of the bare EE $W(\delta=0)$ is linearly proportional to $\delta$.} 
\label{fig_wd_scaling}
\end{figure}

Now, we can investigate the high-frequency behavior of EE which shows a plateau like nature and the saturation value 
is again determined by the $\delta$ (see Fig.~(\ref{fig_wd_scaling})). The high-frequency behavior is simply explained by taking into consideration the fact that the 
periodically
varying transverse field vanishes on average. One can also use Magnus expansion to probe this limit \ct{blanes09}. Therefore, the effective environmental
Hamiltonian in this high-frequency limit is given by 
\be
{\cal H}^{\rm eff}_k(\pm)= (-1\pm \delta +\cos k)\sigma_z+\sin k \sigma_x
\label{eq_heff}
\ee
The effective quasi-energy when $\delta < 1$ is given by $\mu^{\rm eff}_k(\pm)\simeq\pm2(1\pm\delta/2)\sin (k/2)$. Therefore,
one can naively conclude by continuing the analogy of  effective quasi-energy to the EE that the deviation in 
EE from the uncoupled case is proportional to $\delta$ (see the inset of Fig.~(\ref{fig_wd_scaling})).  The EE for the positive channel saturates at a higher
value than that of the negative channel.

One can exactly show that $W(\pm)=W(\delta=0)\pm O(\delta)$ in the high frequency limit where the dynamics of the system is governed by the critical Hamiltonian with 
$\delta$ (\ref{eq_heff}). The quasi-states in this limit are given by $\ket{\eta_k^{+}(\pm)}=(\cos(\theta_\pm/2),~\sin(\theta_\pm/2))^{\cal T}$ and 
$\ket{\eta_k^{-}(\pm)}=(-\sin(\theta_\pm/2),~\cos(\theta_\pm/2))^{\cal T}$ where $\theta_\pm={\rm arctan}[\sin k/(-1\pm \delta +\cos k)]$ and ${\cal T}$ denotes the transpose of a 
matrix. The initial ground state wave-function can be written as $\ket{\phi_{g,k}}=(\cos( \theta_g/2),~ \sin(\theta_g/2))^{\cal T}$ where
$\theta_g={\rm arctan}[\sin k/(-2 +\cos k)]$. $c_k^+(\pm)=\cos(\theta_g/2-\theta_\pm/2)$ and $c_k^-(\pm)=\sin(\theta_g/2-\theta_\pm/2)$.
The work done for the two channels can be simplified to the following form
\ba
W(\pm)&=&(1/N)\sum_k\biggl[ \biggl( \sin k \sin \theta_\pm + (-1\pm\delta+\cos k)\cos \theta_\pm \biggr)\non \\
&\times&  \cos\biggl( \theta_g-\theta_\pm \biggr)  +\sqrt{(2\pm\delta+\cos k)^2+\sin^2 k} \biggr]
\label{eq_wds}
\ea
Now, one can expand the $\theta_\pm$ in the power series of $\delta$ with the constraint that $\delta<1$;
$\theta_\pm=\theta_0 \pm\delta ({\partial \theta_\pm}/{\partial \delta})|_{\delta=0}+\cdots$. The difference between 
two Bogoliubov angles $\theta_g$ and $\theta_\pm$ is given by $\theta_g-\theta_\pm=\theta_S \mp \delta \theta_D$; 
$\theta_S=\theta_g-\theta_0$ and $\theta_D=({\partial \theta_\pm}/{\partial \delta})|_{\delta=0}$. We shall use 
the following assumption to obtain an approximated expression of $W(\pm)$ in the high frequency limit:
$\cos(x+\delta y)=\cos x-\delta y \sin x$. The work done for two channels is then given by
\ba
&W(\pm)&=(1/N)\sum_k \biggl[ \cos \theta_S \cos(k-\theta_0)-\cos \theta_S \cos\theta_0\non \\
&\pm&{\delta \over 2}\biggl(2 \cos \theta_0\cos\theta_S+4\theta_D\cos ({k \over 2}+\theta_S-\theta_0) \non \\
&\times& \sin {k \over 2}\biggr)  + e_{g,k}(\delta=0)\pm\frac{\delta(2+\cos k)}{e_{g,k}(\delta=0)}\biggr]\non \\
&=& W(\delta=0) \pm O(\delta)
\label{eq_wdf}
\ea

Therefore, we can see that the coupling to the qubit has an appreciable effect in EEs over all the 
three regions of frequency.

\begin{figure}[ht]
\begin{center}
\includegraphics[height=6.0cm]{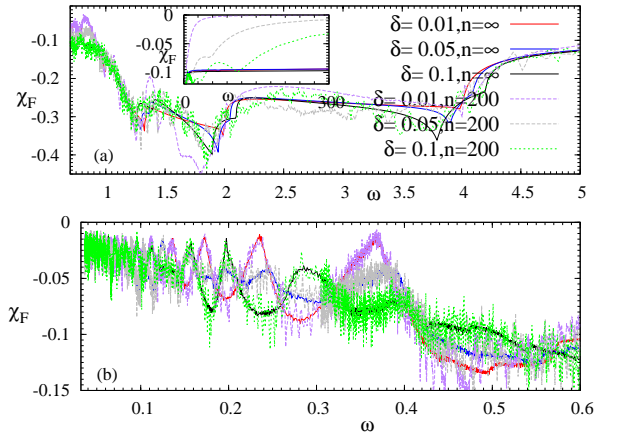}
\end{center}
\caption{Plot shows the variation of DFSS $\chi_F$ for the finite $n$, obtained numerically, and infinite $n$, obtained analytically,
as a function of $\omega$. 
The intermediate and low frequency behavior of $\chi_F$ are shown in Fig. (a) and Fig. (b), respectively.
Fig.~(a) shows that the intermediate frequency peak-dip structure
is more prominently visible in $\chi_F(\infty)$ due to lack of oscillations. Inset of Fig. (a)
shows that the high frequency saturation value of $\chi_F(n)$ is higher as compared to the $\chi_F(\infty)$.
Fig.~(b) depicts that the $\chi_F(n)$ is in very good agreement with the $\chi_F(\infty)$ in the low frequency regime.} 
\label{fig_df1}
\end{figure}

We shall now focus on the behavior of DFSS as a function of frequency. We first present a  comparative study between the two quantities 
$\chi_F(n)$ (\ref{eq_df1}) obtained numerically  for a large value of $n$ and $\chi_F(n\to \infty)$ (\ref{eq_df3}) obtained analytically 
 with frequency. In Fig.(\ref{fig_df1}(a) and (b)), one can observe that the $\chi_F(\infty)$ is matching closely with $\chi_F(200)$ in intermediate and low 
frequency regime except the fact that 
 DFSS for any finite $n$ oscillates rapidly around the mean curve designated by $\chi_F(\infty)$. 
%By investigating Eq.~(\ref{eq_df1}) and
% Eq.~(\ref{eq_df3}), one can see that $\chi_F(n)$ always has a larger value than that of the $\chi_F(\infty)$; this feature  is  
% clearly visible in the high frequency limit due to the oscillation free smooth behavior of the $\chi_F(n)$ (see the inset of Fiq.~(\ref{fig_df1})).
% It is also evident from the Fig. (\ref{fig_df1})  that
% the $\chi_F(\infty)$ gives a better fit for
% $\chi_F(n)$ when $\delta$ becomes sufficiently small i.e., $\delta\ll 1$; the smaller value of $\delta$ allows us to negelect the cross terms in Eq.~(\ref{eq_df3}) which, otherwise, 
% cause a small amount of deviation in a narrow frequency interval ($0.6<\omega<0.9$) for $\delta<1$ if one compares $\chi_F(n)$ with $\chi_F(\infty)$. 
In this low frequency 
regime, $\log(D_k^{\rm dec})$ is maximally contributing to the integral of $\chi_F(\infty)$ where as in the high frequency regime $x'_1$ significantly contributes 
to the $\chi_F(\infty)$. 
%Fig. (\ref{fig_df1}(b)) shows that the $\chi_F(\infty)$ follows the $\chi_F(n)$ with some rapid oscillations.
%On the other hand, 
The high frequency behavior of numerically obtained $\chi_F(n)$ is depicted in the inset of Fig. (\ref{fig_df1}(a)) showing the deviation from
the approximated expression of 
$\chi_F(\infty)$ (\ref{eq_df3}) where ``cross terms" $x'^a_1 x'^2_m$ (with $m\neq 1$) are neglected.
%Therefore, the high frequency saturation value which is dependent on $\delta$ can not be accurately estimated from $\chi_F(\infty)$ (\ref{eq_df3})
%obtained after neglecting the cross term of $O(\delta^2)$  arising from the product $x'^a_1 x'^2_m$ with $m\neq 1$.
Furthermore, in the intermediate frequency regime shown in Fig.~(\ref{fig_df1}(a)), the analytic expression $\chi_F(\infty)$
is sufficient to quantify the behavior of $\chi_F(n)$ as the ``cross term" becomes very small;
the decohered  part and the $x'_m$ part both contribute to the $\chi_F(\infty)$.

\begin{figure}[h!]
\centering
\subfigure{
\includegraphics[width=4.1cm,height=4.3cm]{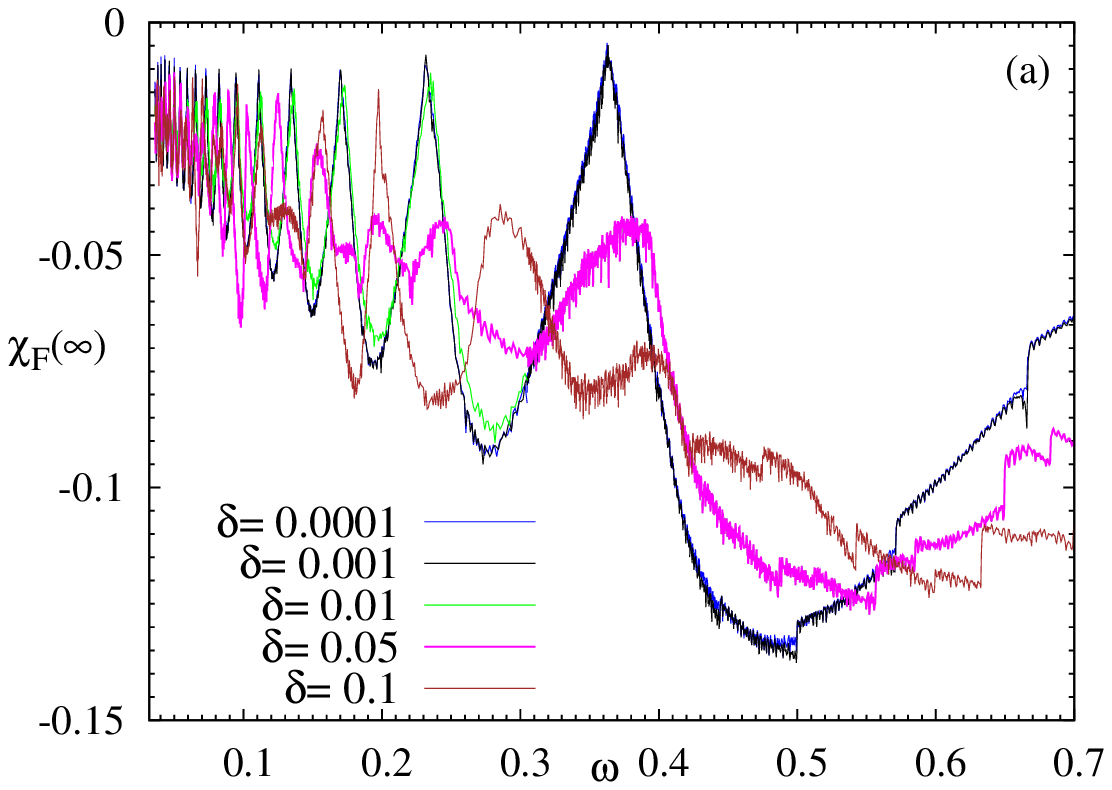}}
\subfigure{
\includegraphics[width=4.1cm,height=4.3cm]{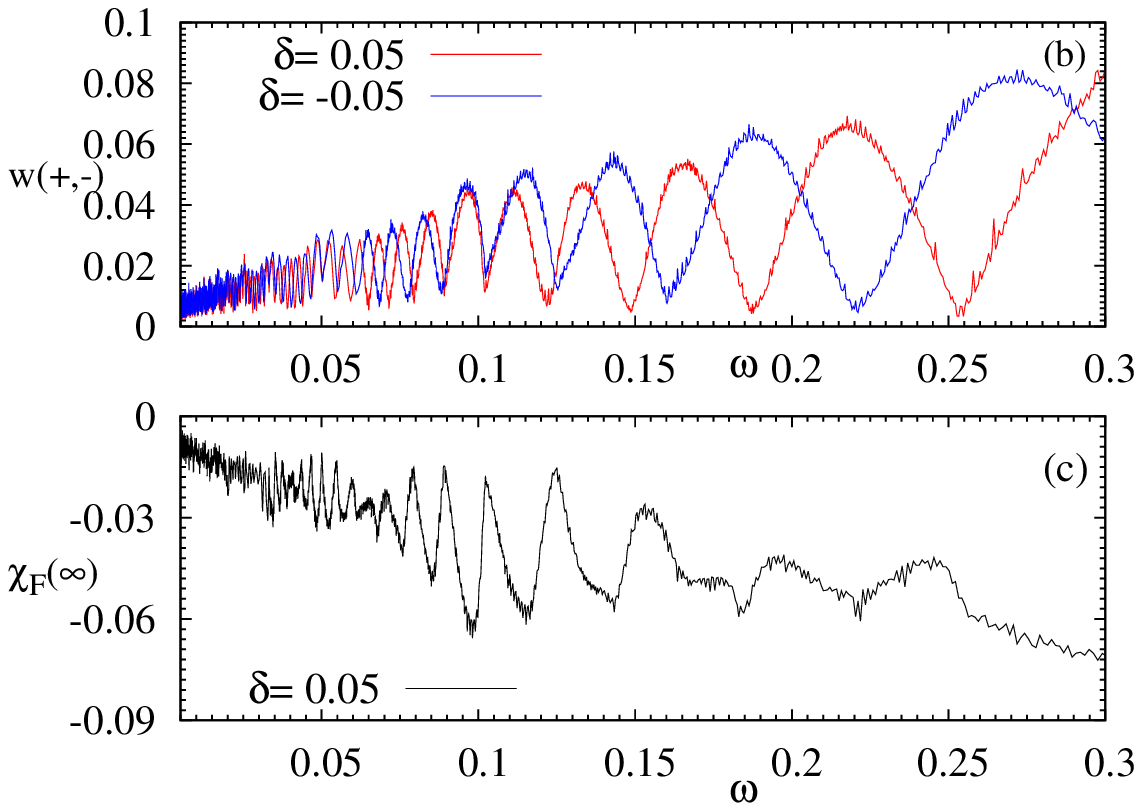}}
\caption{The behavior of $\chi_F(\infty)$ is strongly dependent on $\delta$ as shown in Fig.~(a). 
Fig.~(b) shows the variation of EE for the positive and negative channels with $(h\pm\delta)$ in the low $\omega$ limit. 
Fig.~(c) depicts the response of $\chi_F(\infty)$  as a function of frequency.} 
 \label{fig:DFSS_lf}
 \end{figure}

Now, we shall concentrate on the behavior of  DFSS $\chi_F(\infty)$  as a function of frequency. We also present a comparative study between $\chi_F(\infty)$ and the 
EEs $W(\pm)$. Here as well we will study the $\chi_F(\infty)$ in three different frequency regimes.
In Fig.~(\ref{fig:DFSS_lf}(a)), we study the low-frequency characteristics of
DFSS which  qualitatively shows similar type of oscillations as observed in the case of $W(\pm)$. One can fully understand the 
behavior of $\chi_F(\infty)$  shown in Fig.~(\ref{fig:DFSS_lf}(a))
by comparing the
Fig.~(\ref{fig:DFSS_lf}(b)) with the  Fig.~(\ref{fig:DFSS_lf}(c)).  We can immediately conclude that the behavior of two channels of EEs
actually determine the behavior of DFSS. We can see that when the oscillations for $W(+)$ coincides with that of the $W(-)$ we get nice oscillations in DFSS as an
outcome of constructive superposition between the contribution coming from $\ket{\phi_k(+)}$ and $\ket{\phi_k(-)}$; this 
constructive interference occurs when the oscillations for $J_0((2+4\delta)/\omega)$ matches with that of the $J_0((2-4\delta)/\omega)$. 
On the other hand, DFSS also shows relatively flat regions as an effect
of destructive interference between the two channels of the environmental wave-function. The interplay between two frequencies of
oscillations leads to a  beating like pattern of DFSS. Figure~(\ref{fig:DFSS_lf}(a)) leads to the observation that the beating is more 
prominently visible for relatively higher values $\delta$. On the other hand, beating like pattern gradually turn into simple oscillatory 
behavior, governed by $J_0(2/\omega)$, as one decreases $\delta$.

\begin{figure}[ht]
\begin{center}
\includegraphics[height=6.0cm]{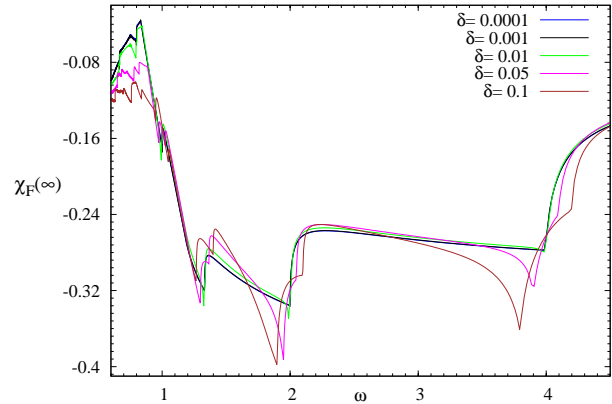}
\end{center}
\caption{Plot shows that $\chi_F(\infty)$ exhibits a qualitatively similar behavior of that of the $W(\pm)$; although, 
the dip positions are dependent on $\delta$.} 
\label{fig:df_if}
\end{figure}

In the intermediate frequency range, DFSS shows dip at the resonance frequency $\omega_r(\pm)$ (see Fig.~(\ref{fig:df_if})). The wave-functions for the two channels
correspond to two different sets of $\omega_r$  where the resonance happens for that particular channel. As a consequence the DF being the 
modulus square overlap of the two wave-function evolving through two channels, $\Pi_k|\langle \phi_k(+)|\phi_k(-)\rangle|^2$, exhibits a change in its behavior at those
$\omega_r(\pm)$; and this is reflected in the response of DFSS at this intermediate frequency range. The behavior of DFSS around $\omega_r(\pm)$ is more prominently visible for 
higher values of $\delta<1$. One can note that DFSS exhibits a dip at those resonance frequencies where 
$W(\pm)$ shows a peak. This is due to the fact that at those frequencies the wave-functions associated with the two channels
become maximally deviated from each other.

\begin{figure}[h!]
\centering
\includegraphics[width=7.5cm,height=4.3cm]{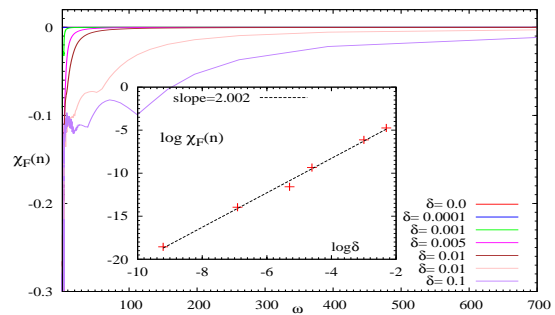}
\caption{Plot shows high frequency saturation behavior of $\chi_F(n)$. Inset shows that saturation value of $\chi_F(n)$ is deviated by an amount 
$\delta^2$ from the bare $\chi_F(n)$ with coupling strength $\delta=0$.} 
 \label{fig:df_scaling}
 \end{figure}

% \begin{figure}[ht]
% \begin{center}
% \includegraphics[height=6.0cm]{figures/df_hf_scaling.eps}
% \end{center}
% \caption{DFSS shows high frequency saturation behavior. Inset shows that saturation value of DFSS is deviated by an amout 
% $\delta^2$ from the bare DFSS with coupling strength $\delta=0$.} 
% \label{fig:df_scaling}
% \end{figure} 

At the end, we shall probe the high frequency behavior of DFSS. To observe the saturation value of DFSS, $\chi_F(\infty)$ is not an 
accurate quantity to be studied. This is due to the fact that high frequency saturation value of $\chi_F(\infty)$ does not 
tend towards zero when $\delta\to 0$ and vanishes for $\delta=0$. One can numerically check that in the high frequency regime, 
``cross term" $x'^a_1x'^2_m$ with  $m\neq1$, contributes significantly to $\chi_F(\infty)$. This
``cross term" gives an $O(\delta^2)$ correction. Therefore, 
$\chi_F(\infty)$ (\ref{eq_df3}) is not an accurate expression which  can correctly describe the behavior of the high frequency
saturation value of DFSS as a function of $\delta$. 
%By analysing $\chi_F(\infty)$ (\ref{eq_df3}) with $\delta=0$, one can see that momemtum sum of $\log(D_k^{\rm dec})$ does not 
%cancel with the other momemtum sum involving logarithm of $x'_1$ term.
One can infer by observing Fig. (\ref{fig:df_scaling}) that the $\chi_F(n)$ is the appropriate quantity 
to examine the saturation behavior with $\delta$ in high frequency limit as the absolute value of $\chi_F(n)$ becomes smaller as one decreases 
$\delta$ and zero if $\delta=0$. Inset of Fig. (\ref{fig:df_scaling}) shows that the saturation value 
is proportional to $\delta^2$. This observation of vanishing $\chi_F(n)$  can be 
justified by analyzing the Eq.(\ref{eq_df1}) with a comparative study between the decohered value $D_k^{\rm dec}$ and the $T'_m$ terms.
One can see that $\delta=0$ leads to the following fact: $D^{\rm dec}_k$ and the cross terms consisting of $x_m\cos\alpha_m$'s  sum up to
unity for each momentum mode while $n$ has a finite value; $\cos\alpha_m=1$ as $\delta=0$. This leads to the 
observation of a vanishing DFSS $\chi_F(\delta=0,n)$ in all frequencies. Therefore, we can infer that $\chi_F(\infty)$ follows the behavior
of $\chi_F(n)$ in all the other frequency regimes  except in the high frequency regime.

We shall now concentrate on the functional behavior of $\delta$ over the high frequency saturation value of DFSS as shown in Fig.~(\ref{fig:df_scaling}). As we mention previously for the case of EEs, one can get
a qualitatively approximate expression of
the quasi-states and
the quasi-energies by diagonalizing the effective Hamiltonian in Eq.~(\ref{eq_heff}). We can therefore define an effective expression of the DF, qualitatively valid only for infinite frequency regime, in the momentum space
 is given by
\ba
D^{\rm eff}_k&=&| \langle\eta_k(+)|\eta_k(-)\rangle|^2=\cos^2\biggl({\theta_+-\theta_-\over 2}\biggr)\nonumber \\
&\simeq& 1-{\delta^2 \over 4} \sin^2{k\over 2}
\label{eq_lf}
\ea
 DF is $D^{\rm eff}=\Pi_{k>0}D^{\rm eff}_k$. This over simplified expression of $D^{\rm eff}_k$
is not a quantitatively accurate expression because it does not give the  high frequency saturation value of DFSS $\chi_F(n)$ correctly.
We can only infer from the above
expression that the infinite frequency saturation value of DFSS (\ref{eq_df1}) of the qubit $\chi_F(n)$ is deviated
from that of the  $\chi_F(n,\delta=0)$ by an amount proportional to $\delta^2$.

One can also use the usual definition of fidelity 
susceptibility \ct{zanardi06} $\chi_F(n)=\log(D(n))/N=\delta^2\chi_{\cal F}(n)$ to probe its high frequency saturation behavior where $\chi_{\cal F}(n)$ in our case is given by
\ba
\chi _{\cal F}(n) &=&\biggl[\left\langle \frac{\partial }{\partial \delta } \phi_k(+,nT)
\left\vert \frac{\partial }{\partial \delta } \phi_k(-,nT)\right.
\right\rangle  \non\\
&&-\left\langle \left. \frac{\partial }{\partial \delta } \phi_k(+,nT)
\right\vert  \phi_k(-,nT)\right\rangle \non\\
&&\left\langle  \phi_k(+,nT)\left\vert
\frac{\partial }{\partial \delta } \phi_k(-,nT)\right. \right\rangle\biggr]_{\delta=0} .
\label{eq_fs}
\ea
Therefore, it is evident from the above Eq.~(\ref{eq_fs}) that the $\chi_F(n)\propto \delta^2$ as 
$\chi_{\cal F}(n)$ is independent of $\delta$. We can see that in all the frequency regime the  EE and DFSS are modified by the small 
parameter $\delta$ and their behavior are connected to each other 
through this small parameter $\delta$.

% In the high frequency limit one can examine the behavior of $\chi_F(\infty)$ with $\delta$. The decohered part 
% $\log D_k^{\rm dec}$ of $\chi_F(\infty)$ contributes in the $\delta^2$ order where the other part , consisting of $x'_m$ and $y'_m$,
% contribute in $O(\delta)$. The terms with the exponential phase factor has both $O(\delta)$ and $O(\delta^2)$ terms in it.  As a result, $x'_2, ~x'_3,~x'_4,~x'_5$ yield 
% first order in $\delta$ terms and $x'_1$ and $x'_6$ gives second order term in $\delta$. It is therefore possible to conclude that in the 
% high frequency limit the $\chi_F(\infty)$ behaves linealy on $\delta$. It is not straight forward to obtain an approximated expression of
% $\chi_F(\infty)$ in the form of $\chi_F(\infty,\delta)=\chi_F(\infty,\delta=0)+O(\delta)$. This is due to the fact that 
% $\chi_F(\infty,\delta=0)=0$ and Consequently terms with $\delta$ do not cause a correction to the bare value rather completely modify 
% $\chi_F(\infty,\delta$) giving a finite value. 

\section{conclusion}
\label{remarks}
In conclusion, we choose a central spin model where a single qubit is globally coupled to an environmental Ising spin chain periodically driven across its QCP 
and numerically study the EE associated with it
as a function of frequency in the infinite time limit; we also numerically
investigate  the DFSS of the qubit after a large number of period with frequency. 
Our aim is to characterize the behavior of EE and DFSS when a small system-environment coupling parameter is present and   analyze 
 their behavior by plausible analytical argument. 
 The coupling to the qubit gives rise to two channels of time 
 evolution for the environment and  consequently leads to two separate species of Floquet operators.
In the process, we show that the coupling strength $\delta$
can influence the position of quasi-degenerate momentum mode in the Floquet spectrum associated with the two species of Floquet operators. We have
two sets of EEs $W(\pm)$ each of them is associated with the one of these channels. We show that the low-frequency oscillations of $W(\pm)$ for two channels  are dominated by the two different
Bessel functions with argument dependent on $\delta$. The position of the resonance peaks at intermediate frequencies for 
two channels are different from each other due to the finite coupling strength. Finally, high frequency saturation value of EEs for two channels are 
dependent on $\delta$ in such a way that the EE for the positive channel takes a higher value of saturation than that of the negative channel.

% We show that when the coupling strength is small, the behavior of EE and DFSS both as a function of frequency are affected
%by the above. 

In parallel, we find an analytical expression for the $\chi_F(n\to\infty)$
which can successfully predict the behavior of the DFSS $\chi_F(n)$, obtained numerically for a large value of $n$, in the low and intermediate frequency regime.
We show that the behavior of DFSS of the qubit $\chi_F(\infty)$ in the above frequency regions
can be speculated by understanding the behavior of EEs for the two channels of evolution associated with environmental spin chain. 
DFSS exhibits low-frequency beating like pattern originated from the interplay between
two Bessel functions, governing the oscillations of $W(\pm)$, with two different arguments. $\chi_F(\infty)$ displays dips at intermediate frequencies where the $W(\pm)$ show peaks. The position of these 
dips, appearing due to   destructive interference between two channels of environment, are dependent on $\delta$. At the end, the DFSS tends to 
saturation value at high frequency which is correctly quantified by $\chi_F(n)$ instead of $\chi_F(\infty)$. The saturation value of $\chi_F(n)$ is deviated from the bare saturation value of DFSS $\chi_F(n,\delta=0)=0$ by an amount 
proportional to $\delta^2$. This deviation of $O(\delta^2)$ can be estimated from the effective DFSS defined for the high frequency 
static Hamiltonian which also successfully predicts the correction of $O(\delta)$ in EE at the high frequency limit. Therefore, in all the above three frequency region the behavior of DFSS is closely tied with the 
behavior of EEs for two different channels.

Moreover, the Loschmidt echo has been experimentally investigated  in an antiferromagnetic Ising spin chain with finite number of spins using NMR quantum simulator \ct{zhang09}. 
The  periodic driving has also been experienced experimentally leading to many interesting observations \ct{kitagawa12}. Therefore, experimental verification of our work might be possible 
by employing a time periodic model in the large scale NMR quantum simulator.

\section{Acknowledgements}
The author acknowledges Amit Dutta for critically reading the manuscript and giving useful suggestions. The author also thanks Victor 
Mukherjee for valuable discussions.
%%%% Appendix %%%%%%%
\appendix
\section{Exact Calculation of $\chi_F(\infty)$}
\label{ch_append}

In this appendix, we shall explicitly calculate the DFSS $\chi_F(\infty,\delta)$ (\ref{eq_df3}) in the $n\to \infty$ limit with $\delta<1$. We shall 
calculate each term of $D(n)$ including $T_m$'s and show their functional dependence on $\delta$. 
The quasi-state $\ket{\eta_k^{\pm}(\pm)}$, associated with the two species of Floquet operators ${\cal F}_k(\pm)$, can be expanded in the powers of $\delta$ 
\be
\ket{\eta^{\pm}_k(\pm)}=\ket{\eta^{\pm}_k(0)} \pm  \delta~\frac{\partial \ket{\eta_k^{\pm}(\pm)}}{\partial \delta}\vert_{\delta=0} +\cdots,
\label{eq_a1}
\ee
where $\ket{\eta^{\pm}_k(0)}$ is the quasi-state of bare Floquet operator ${\cal F}_k(\delta=0)$. Now, one can easily calculate the 
 the Floquet coefficients $c^{\pm}_k(\pm)$ in the power series of $\delta$
\be
c^{\pm}_k(\pm)=c^{\pm}_k(0)\pm  \delta~\frac{\partial c_k^{\pm}(\pm)}{\partial \delta}|_{\delta=0} +\cdots.
\label{eq_a2}
\ee
Here, $({\partial c_k^{\pm}(\pm)}/{\partial \delta})=\langle\phi_{g,k}|{\partial \ket{\eta_k^{\pm}(\pm)}}/{\partial \delta}$. Similarly,
the overlap $\langle \eta^{\pm}_k(+)|\eta^{\pm}_k(-)\rangle$ can also be expanded in the ascending powers of $\delta$. 
Now, $\langle \eta^{\pm}_k(+)|\eta^{\pm}_k(-)\rangle$ is given by 
\be
\langle \eta^{\pm}_k(+)|\eta^{\pm}_k(-)\rangle= 1 -\delta^2 ~
\biggl[ \frac{\partial\langle \eta^{\pm}_k(+)|\partial|\eta^{\pm}_k(-)\rangle}{\partial \delta} \biggr]_{\delta=0} 
\label{eq_a3}
\ee
On the other hand, $\langle \eta^{\pm}_k(+)|\eta^{\mp}_k(-)\rangle$ is given by 
\ba
\langle \eta^{\pm}_k(+)|\eta^{\mp}_k(-)\rangle&=&-\delta~
\biggl[ \frac{\langle \eta^{\pm}_k(0)|\partial|\eta^{\mp}_k(-)\rangle}{\partial \delta} \biggr]_{\delta=0}\non \\
&+&\delta\biggl[ \frac{ \partial \langle \eta^{\pm}_k(+)|\eta^{\mp}_k(0)\rangle}{\partial \delta} \biggr]_{\delta=0} + O(\delta^2) \non
\label{eq_a4}
\ea

Now, we shall analyze each 
term of $D(n)$ and its dependence on  $\delta$.

\ba
&&|c_k^+(+)|^2|c_k^+(-)|^2 ~|\langle\eta_k^+(+)|\eta_k^+(-)\rangle|^2 \to |c_k^+(0)|^4 + O(\delta^2)\non \\
&&|c_k^-(+)|^2|c_k^-(-)|^2 ~|\langle\eta_k^-(+)|\eta_k^-(-)\rangle|^2 \to |c_k^-(0)|^4 + O(\delta^2)\non \\
&&|c_k^+(+)|^2|c_k^-(-)|^2|\langle\eta_k^+(+)|\eta_k^-(-)\rangle|^2 \to O(\delta^2) \non \\
&&|c_k^-(+)|^2|c_k^+(-)|^2 |\langle\eta_k^-(+)|\eta_k^+(-)\rangle|^2 \to O(\delta^2) \non \\
&& T'_1 \to |c_k^+(0)|^2 |c_k^-(0)|^2 + O(\delta^2),~T'_2 \to O(\delta)\non \\
&& T'_3 \to O(\delta), ~T'_4 \to O(\delta), T'_5 \to O(\delta), ~T'_6 \to O(\delta^2) \non 
\label{eq_a5}
\ea
One can probe the $\delta$-dependence on the $x'_m$ using the above observation. This leads to the following fact
$x'_1\to O(\delta^0)+O(\delta^2)+O(\delta^4)$, $x'_m\to O(\delta)+O(\delta^3)$ with $m=2,3,4,~5$ and $x'_6\to O(\delta^2)+O(\delta^4)$. 

Therefore, one can write an approximated expression of $\chi_F(n\to \infty)$ by a momentum space integration over the $x'_m\cos\alpha_m$. 
In order to estimate $\chi_F(n\to \infty)$, we define a quantity $A$ which is given by
\ba
&A&=\log\biggl[1+\sum_{m=1}^6 z_m\biggr]\nonumber\\
&=& \sum_{m} z_m -{1 \over 2} \sum_{m,n} z_m z_n+ {1 \over 3} \sum_{m,n,p} z_m z_n z_p +\cdots 
\label{eq_a6}
\ea
$\chi_F(n)$  is then given by $\chi_F(n)={1 \over N}\sum_k [ \log(D_k^{\rm dec})+A]$.
Here, $z_m={x'_m  }\cos\alpha_m$. $x'_m$ and $\alpha_m$ are $k$ dependent functions.

One can obtain the following type of  terms by 
decomposing the above Eq.~(\ref{eq_a6}).
\ba
z_i&=&\sum_i x'_i \cos \alpha_i,~ z_i z_j= \sum_{i,j}x'_i x'_j \cos \alpha_i \cos \alpha_j\nonumber\\
&=& {1 \over 2}\sum_{i,j}x'_i x'_j \biggl(\cos(\alpha_i+\alpha_j)+\cos(\alpha_i-\alpha_j)\biggr),~\nonumber\\
z_i z_j z_k&=&  \sum_{i,j,k} x'_i x'_j x'_k \cos \alpha_i \cos \alpha_j \cos \alpha_k\nonumber\\
&=& {1 \over 4}\sum_{i,j,k} x'_i x'_j x'_k \biggl( \cos(\alpha_i+\alpha_j+\alpha_k)+\cos(\alpha_i+\alpha_j-\alpha_k)\nonumber\\
&+&\cos(\alpha_i-\alpha_j+\alpha_k)+\cos(\alpha_i-\alpha_j-\alpha_k) \biggr)\nonumber\\
z_i z_j z_k z_l&=& \sum_{i,j,k,l} x'_i x'_j x'_k x'_l \cos \alpha_i \cos \alpha_j \cos \alpha_k \cos \alpha_l\nonumber\\
&=&{1 \over 8}\sum_{i,j,k,l} x'_i x'_j x'_k x'_l \biggl( \cos(\alpha_i+\alpha_j+\alpha_k +\alpha_l)  \nonumber\\
&+&\cos(\alpha_i-\alpha_j+\alpha_k+\alpha_l)+\cos(\alpha_i+\alpha_j-\alpha_k +\alpha_l) \nonumber\\
&+&  \cos(\alpha_i-\alpha_j-\alpha_k +\alpha_l)  +\cos(\alpha_i+\alpha_j+\alpha_k -\alpha_l)\nonumber\\
&+& \cos(\alpha_i-\alpha_j+\alpha_k -\alpha_l)+ \cos(\alpha_i+\alpha_j-\alpha_k -\alpha_l)\nonumber\\
&+& \cos(\alpha_i-\alpha_j-\alpha_k -\alpha_l) \biggr)\non
\ea
Here, $\alpha_i$ is a function of $\mu^{\pm}_k(\pm) nT$. As $n\to \infty$, the complete momentum integration over 
the power series of $z_i$
 gives non-zero values when the argument of the cosine term vanishes. Now, the argument of the cosine term vanishes only when an even number of
sum is involved inside the argument with alternating signs but the same index. For example, the integration over $z_i z_j$ yields $\sum_i x'^2_i/2$ as $\cos(\alpha_i+\alpha_j)$ does not 
survive under integration while $\cos(\alpha_i-\alpha_j)$ survives only for $i=j$. Similarly, integration over $z_i z_j z_k z_l$ and 
$z_i z_j z_k z_l z_m z_n$ contribute
$\sum_i 3 x'^4_i/8$ and $\sum_i 5 x'^6_i/16$ to the $\chi_F(\infty)$, respectively.

One can see that $\chi_F(\infty)$ receives a contribution of $O(\delta^0)$ with $O(\delta^2)$ from 
the decohered part. Now, for $m\neq 1$ and $m \neq 6$ the leading order term from the non-decohered part, coming from $x'^2_m$, is $O(\delta^2)$.
The next leading order term of $O(\delta^4)$ is therefore coming from $x'^4_m$; $x'^n_m\to O(\delta^n)$. For $m=6$, 
$x'^n_6\to O(\delta^{2n})$. In the case of $m=1$, the  $x'^{2p}_1$ ($p$ is an integer) generates all even power terms in $\delta$
starting from 
$O(\delta^0)$ to $O(\delta^{2p})$.  Therefore, in order to incorporate the complete $O(\delta^0)$ contribution, 
one has to consider the full series consisting the even powers in $x'_1$; this results in the closed form expression 
$\log(2/(1+\sqrt{1-x'^2_1}))$. The others power series can be safely truncated by retaining only the leading contributions. One can exclude $x'^2_6$ term as it has 
$O(\delta^4)$ correction.

%On the other hand, integration over $\gamma$  gives a similar series of $y'_i$; $\gamma^2 \rightarrow \sum_i y'^2_i/2$ and $\gamma^4 \rightarrow \sum_i 3 y'^4_i/8$.
%In this even power series of $x'_m$ and $y'_m$, the first order term is 

Here, one can also  get a finite  contribution from even order term like $x'^{2p}_i x'^{2q}_j$, referred as the ``cross term" in Eq.~(\ref{eq_df3}),
 with $i\neq j$, $p$ and $q$ can be any positive integer.
 This type of terms appear when the argument of the cosine term
vanishes in a pair. For example, $z_i z_j z_k z_l$ yields $x'^2_i x'^2_j$ type of terms from the momentum integration over 
$\cos(\alpha_i-\alpha_l+\alpha_j -\alpha_k)$ with $i=l$ and $j=k$. 
This type of fourth order terms  are the lowest order term from which the series of $x'^{2p}_i x'^{2q}_j$ 
 starts contributing to the momentum integral of $\chi_F(n \to \infty)$. But, the 
contribution coming from this term is $O(\delta^4)$ except the product terms like $x'^{2p}_1 x'^2_m$ (with $m\neq 1,6$) which has an $O(\delta^2)$ correction.
A closed form expression can not be obtained for this $O(\delta^2)$ terms. 
%But, the coefficient of this product term of $O(\delta^2)$ is 
%very small as compared to the coefficient $O(\delta^2)$ obtained for $\log(2/(1+\sqrt{1-x'^2_1}))$.
Numerical investigation suggests that the product $x'_1 x'_m$ becomes insignificantly small in low and intermediate frequency. Therefore, 
one can neglect the above $O(\delta^2)$ contribution in low and intermediate frequency regime; $\chi_F(\infty)$ can be constructed by keeping only the leading order contribution 
coming from the decohered part and $x'^{2p}_m$ (with $m=1,~\cdots,~6$) part. Combining all the significant contribution  one can write the approximated of $\chi_F(\infty)$ (\ref{eq_df3}) 
in low and intermediate  frequency regime.

\end{document}